# Pure Nuclear Fusion Bomb Propulsion

F.Winterberg

University of Nevada, Reno

March 2008

# **Abstract**

Recent progress towards the non-fission ignition of thermonuclear micro-explosions raises the prospect for a revival of the nuclear bomb propulsion idea, both for the fast transport of large payloads within the solar system and the launch into earth orbit without the release of fission products into the atmosphere. To reach this goal three areas of research are of importance:

1) Compact thermonuclear ignition drivers.

2) Fast ignition and deuterium burn.

3) Space-craft architecture involving magnetic insulation and GeV electrostatic potentials.

# 1. Introduction

With chemical propulsion manned space flight to the moon is barely possible and only with massive multistage rockets. For manned space flight beyond the moon, nuclear propulsion is indispensible. Nuclear thermal propulsion is really not much better than advanced chemical propulsion. Ion propulsion, using a nuclear reactor driving an electric generator, has a much higher specific impulse, but not enough thrust for short interplanetary transit times, needed for manned missions. This leaves nuclear bomb propulsion as the only credible option. Both its thrust and specific impulse are huge in comparison.

For a history of nuclear bomb propulsion, reference is made to a long article by A.R.Martin and A.Bond [1]. Under the name project Orion, it was studied in great detail under the leadership of Theodore Taylor and Freeman Dyson. Its history has been published by George Dyson [2], the son of Freeman Dyson. The project was brought to a sudden halt by the nuclear test ban treaty, motivated by the undesirable release of nuclear fission products into the atmosphere. For pure fusion explosions the situation is much more favorable, because neutron activation of the air is much less serious. In nature it happens all the time by cosmic rays.

A first step in this direction is the non-fission ignition of thermonuclear micro-explosions, expected to be realized in the near future. By staging and propagating thermonuclear burn, it should lead to the non-fission ignition of large thermonuclear explosive devices.

During project Orion detailed engineering studies about bomb propulsion were made. Apart from some basic considerations, the result of this work shall not be repeated here. I rather will focus on three crucially important topics:

1) The architecture of the space craft incorporating the non-fission ignition driver.

2) The fusion explosive.

3) The delivery of the ignition pulse to the fusion explosive.

The launch from the surface of the earth remains the most difficult task. For it a different approach is proposed. The idea to use intense relativistic electron or ion-beam induced nuclear micro-explosions for rocket propulsion was proposed by the author [3,4], and the use of fission-triggered large fusion bomb propulsion for interstellar space flight by Dyson [5]. An electron-beam induced pure nuclear fusion micro-explosion propulsion system (project Daedalus) was extensively studied by the British Interplanetary Society [6].

## 2) The Basic Requirements

As a typical example we consider a space craft with a mass of $M_o = 10^3$ tons $= 10^9$ g, to be accelerated by one $g \cong 10^3$ cm/s$^2$ with a thrust $T = M_o g \cong 10^{12}$ dyn. To establish the magnitude and number of fusion explosions needed to propel the space craft to a velocity of $v = 100$ km/s $= 10^7$ cm/s, we use the rocket equation

$$T = c \frac{dm}{dt} \tag{1}$$

where we set $c \cong 10^8$ cm/s equal to the expansion velocity of the fusion bomb plasma.

We thus have

$$\frac{dm}{dt} = \frac{T}{c} = 10^4 \, g/s = 10 \, kg/s \tag{2}$$

The propulsion power is given by

$$P = \frac{c^2}{2} \frac{dm}{dt} = \frac{c}{2} T \tag{3}$$

in our example it is $P = 5 \times 10^{19}$ erg/s

With $E = 4 \times 10^{16}$ erg equivalent to the explosive energy of one ton of TNT,

P is equivalent to about one nuclear kiloton bomb per second.

From the integrated rocket equation

$$\text{v} = c \ln\left(1 + \frac{\Delta M}{M_o}\right) \qquad (4)$$

where v is the velocity reached by the space craft after having used up all the bombs of mass

$\Delta M$ , one has for $\Delta M << M_o$ ,

$$\Delta M / M_o \cong \text{v}/c \qquad (5)$$

If one bomb explodes per second, its mass according to (2) is $m_o = 10^4$ g.

Assuming that the space craft reaches a velocity of $\text{v} = 100\,\text{km/s} = 10^7$ cm/s, the velocity

needed for fast interplanetary travel, one has $\Delta M = 10^8$ g, requiring $N = \Delta M / m_o = 10^4$

one kiloton fusion bombs, releasing the energy $E_b = 5 \times 10^{19} \times 10^4 = 5 \times 10^{23}$ erg.

By comparison, the kinetic energy of the space craft is $E_s = (1/2) M_o \text{v}^2 = 5 \times 10^{23}$ erg,

about 10 times less. In reality it is still smaller, because a large fraction of the energy released by

the bomb explosions is dissipated into space. If launched from the surface of the earth, one has to

take into account the mass of the air entrained in the fireball. The situation there then resembles

a hot gas driven gun, albeit one of rather poor efficiency. There the velocity gained by the craft

with N explosions, each setting off the energy $E_b$ , is given by

$$\text{v} = \sqrt{2 N E_b / M_o} \qquad (6)$$

For $E_b = 5 \times 10^{19}$ erg, $M_o = 10^9$ g, and setting for $\text{v} = 10\,\text{km/s} = 10^6$ cm/s the escape velocity from earth, one finds that $N \geq 10$. Assuming an efficiency of 10%, about 100 kiloton explosions would there be needed.

One can summarize these estimates by concluding that a very large number of nuclear explosions is needed , which for fission explosions, but also for deuterium-tritium explosions, would become very expensive. This strongly favors deuterium, more difficult to ignite in comparison to a mixture of deuterium with tritium. Here I will try to show how bomb propulsion with deuterium, at worst with a small amount of tritium, might be possible.

## 3. Ways towards the Non-Fission Ignition of Deuterium Nuclear Explosions

With no deuterium-tritium (DT) micro-explosions yet ignited, the ignition of pure deuterium (D) fusion explosions seems to be a tall order. An indirect way to reach this goal is by staging a smaller DT explosion with a larger D explosion. There the driver energy, but not the driver may be rather small. A direct way requires a driver with order of magnitude larger energies.

I claim that the generation of GeV potential wells, made possible with magnetic insulation of conductors levitated in ultrahigh vacuum, has the potential to lead to order of magnitude larger driver energies [7,8]. It is the ultrahigh vacuum of space by which this can be achieved. And if it is the space craft itself, which is charged up to GeV potentials, there is no need for its levitation.

If charged to a positive GeV potential, a gigajoule intense relativistic ion beam below the Alfven current limit can be released from the spacecraft and directed on the D explosive for its ignition. Because the current needed for ignition is below the Alfven limit for ions, the beam is "stiff". The critical Alfven current for protons is $I_A = 3.1 \times 10^7 \beta\gamma [A]$, for GeV protons well in excess of the critical current to entrap the DD fusion reaction products, a condition for detonation [9].

In a possible bomb configuration shown in Fig.1, the liquid (or solid) D explosive has the shape of a long cylinder, placed inside a cylindrical "hohlraum" **h**. A GeV proton beam **I** coming from the left, in entering the hohlraum dissipates part of its energy into a burst of X-rays compressing and igniting the D bomb-cylinder. With its gigajoule energy lasting less than $10^{-7}$ s, the beam power is greater than $10^{16}$ Watt, sufficiently large to ignite the D explosive. The purpose of rocket chamber R containing solid hydrogen will be explained below.

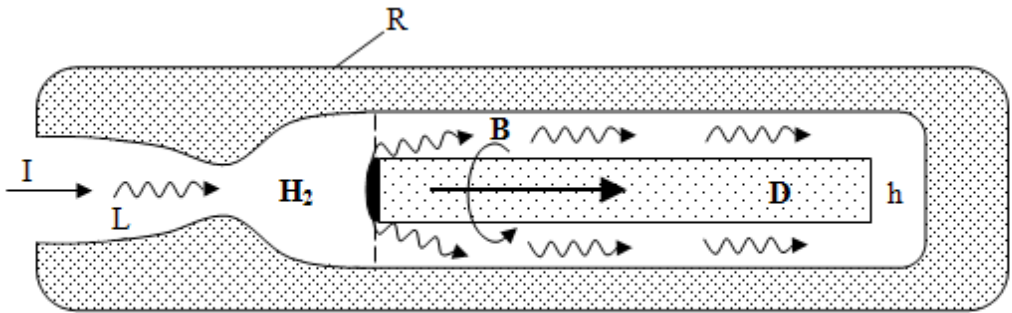

Figure 1

Fig.1: Pure deuterium fusion explosion ignited with an intense ion beam. **D** deuterium rod, **h** hohlraum, **I** ion beam, **B** magnetic field, **R** rocket chamber, **H** solid hydrogen, **L** laser beam to heat hydrogen in rocket chamber.

For a launch from the surface of the earth magnetic insulation inside the earth atmosphere fails, and with it the proposed pure D bomb configuration. Here a different technique is suggested, first proposed in a classified report, dated January of 1970 [10], declassified July 2007. A similar idea was proposed in a classified Los Alamos report, dated November 1970, [11], declassified  July 1979. The idea is to use a replaceable laser for the ignition of each nuclear explosion, with the laser material thereafter becoming part of the propellant. The Los Alamos scientists had proposed to use for this purpose an infrared carbon dioxyde ($CO_2$ ) or chemical laser, but this idea does not work, because the wave length is there too long. In my somewhat earlier report I had suggested to use an ultraviolet Argon ion laser instead. However, since Argon ion lasers driven by an electric discharge have a small efficiency, I had suggested a quite different way for its pumping, illustrated in Fig.2, where the efficiency can be expected to be quite high. As shown in Fig.2, a cylinder of solid Argon is surrounded by a thick cylindrical shell of high explosive, simultaneously detonated from outside, launching a convergent cylindrical shock wave into the Argon. For the high explosive one may choose hexogen with a detonation velocity of 8 km/s. For a convergent cylindrical shock wave the temperature rises as r$^{-0.4}$, where r is the distance from axis of the cylindrical Argon rod. If the shock is launched from a distance of ~1m onto an argon rod with a radius equal to 10cm, the temperature reaches 90 000º K,  just right to excite the upper laser level of Argon. Following its heating to 90 000º K the Argon cylinder radially expands and cools, with the upper  laser level frozen in the Argon. This is similar as in a gas dynamic laser, where the upper laser level is frozen in the gas during its isentropic expansion in a Laval nozzle. To reduce depopulation of the upper laser level during the expansion by super radiance, one may dope to the Argon with a saturable absorber, acting as an "antiknock" additive. In this way megajoule laser pulses can be released within 10 nanoseconds. A laser pulse from a small Q-switched Argon ion laser placed in the space craft can then launch a photon avalanche in the Argon rod, igniting a DT micro-explosion.

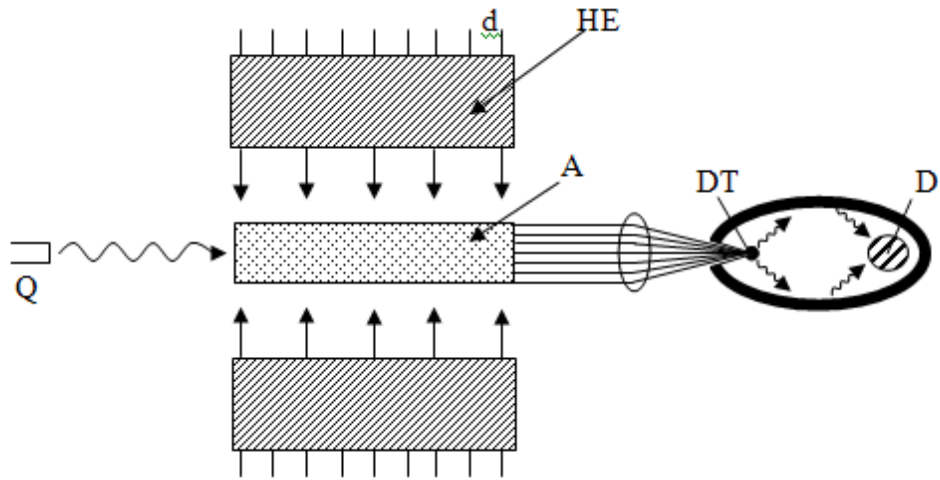

Figure 2

Fig.2: Argon ion laser ignitor, to ignite a staged **DT →D** fusion explosion in a mini-Teller-Ulam configuration. **A** solid argon rod. **HE** cylindrical shell of high explosive, **d** detonators. **Q** Q-switched argon ion laser oszillator.

Employing the Teller-Ulam configuration, by replacing the fission explosive with a DT microexplosion, one can then ignite a much larger D explosion.

As an alternative one may generate a high current linear pinch discharge with a high explosive driven magnetic flux compression generator. If the current I is of the order I=$10^7$ [A], the laser can ignite a DT thermonuclear detonation wave propagating down the high current discharge channel, which in turn can ignite a much larger pure D explosion.

## 4. The Space Craft Architecture

The original idea for the electrostatic energy storage on a magnetically insulated conductor was to charge up to GeV potentials a levitated superconducting ring, with the ring magnetically insulated against breakdown by the magnetic field of a large toroidal current flowing through the ring. It is here proposed to give the space craft a topologically equivalent shape, using the entire space craft for the electrostatic energy storage (see Fig.3). There, toroidal currents flowing azimuthally around the outer shell of the space craft, not only magnetically insulate the spacecraft against the surrounding electron cloud, but the currents also generate a magnetic mirror field which can reflect the plasma of the exploding fusion bomb. In addition, the expanding bomb plasma can induce large currents, and if these currents are directed to flow through magnetic field coils positioned on the upper side of the space craft, electrons from there can be emitted into space surrounding the space craft by thermionic emitters placed on the inner side of these coils in a process called inductive charging [12]. This recharges the space craft for subsequent proton beam ignition pulses. A small high voltage generator driven by a small onboard fission reactor can make for the initial charging, ejecting from the space craft negatively charged pellets.

## 5. Delivery of a GeV Proton Beam onto the Deuterium Fusion Explosive

The space craft is positively charged against an electron cloud surrounding the space craft, and with a magnetic field of the order $10^4$G, easily reached by superconducting currents flowing in an azimuthal direction, the space craft is insulated against the electron cloud up to

GeV potentials. The space craft and its surrounding electron cloud form a virtual diode with a

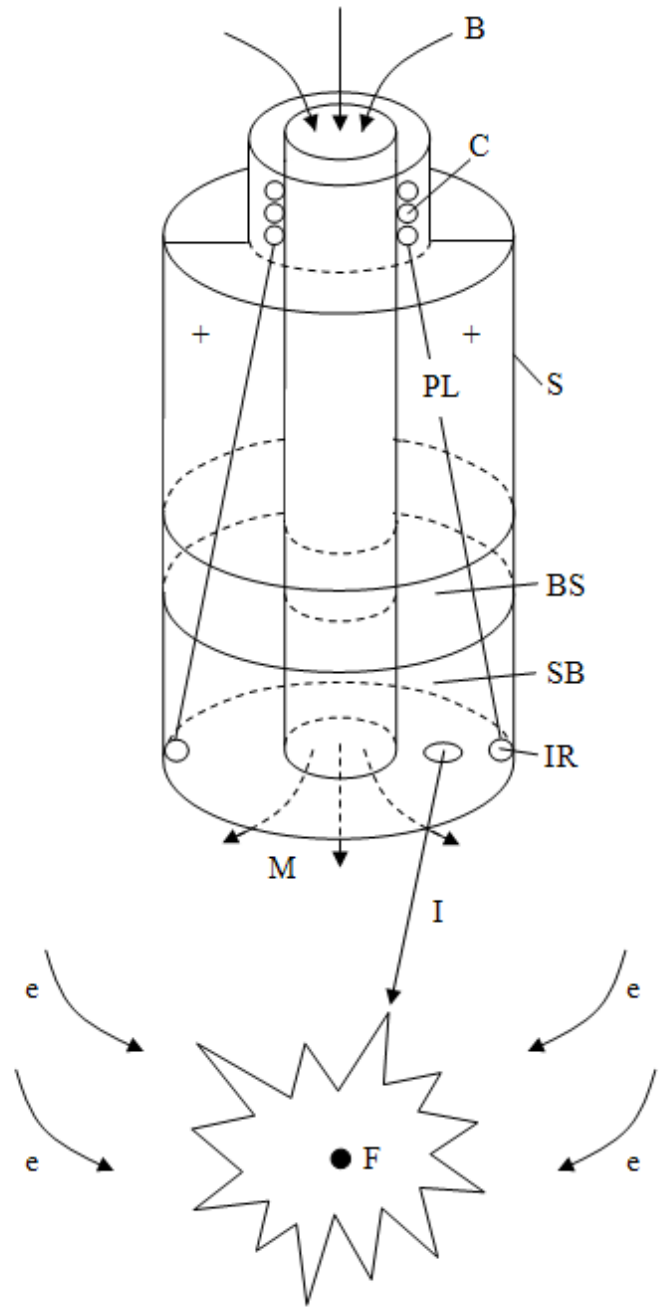

Figure 3

Fig.3: Superconducting "Atomic" Space Ship,   positively charged to GeV potential, with
azimuthal currents and magnetic mirror **M** by magnetic field **B**. **F** fusion minibomb in
position to be ignited by intense ion beam **I**, **SB** storage space for the bombs, **BS**
bioshield for the payload **PL**, **C** coils pulsed by current drawn from induction ring **IR**. **e**
electron flow neutralizing space charge of the fusion explosion plasma.

GeV potential difference. To generate a proton beam, it is proposed to attach a small hydrogen filled rocket chamber **R** to the deuterium bomb at the position where the  proton beam hits the fusion explosive (see Fig.1). A pulsed laser beam from the space craft is shot into the rocket chamber, vaporizing the hydrogen, and emitted as a supersonic plasma jet through the Laval nozzle. If the nozzle is directed towards the space craft, a conducting bridge is established, rich in protons between the space craft and the fusion explosive. Protons in this bridge are then accelerated to GeV energies, hittng the deuterium explosive. Because of the large dimension of the space craft, the jet has to be aimed at the space craft not very accurately.

With the magnetic insulation criterion, $B > E$, ($B, E$, in electrostatic units) where B is the magnetic field surrounding the space craft, then for B $\cong 10^4$G, $E = 3 \times 10^3$ esu $= 9 \times 10^5$ V/cm, one has $E \sim (1/3)B$. A space craft with the dimension $l \sim 10^3$ cm, can then be charged to a potential $El \sim 10^9$ Volts. The stored electrostatic energy is of the order $\varepsilon \sim \left(E^2/8\pi\right) l^3$. For $E = 3 \times 10^3$ esu, and $l = 10^3$ cm, it is of the order of 100 gigajoule. Therefore, only 1% of the electrostatic energy stored is needed for one gigajoule ignition pulse. To discharge not more than one gigajoule, the number of protons in the jet, respectively the amount of hydrogen in the rocket chamber, must be limited. The discharge time is of the order $\tau \sim l/c$, where $c = 3 \times 10^{10}$ cm/s is the velocity of light. In our example we have $\tau \sim 3 \times 10^{-8}$ sec. For a proton energy pulse of one gigajoule the beam power is $3 \times 10^{17}$ erg/s $= 300$ petawatt, large enough to ignite a pure deuterium explosion.

### *References*

[1] A. R. Martin and Alan Bond J. British Interplanetary Society 32, 283-310 (1979)

[2] G. Dyson, Project Orion, Henry Holt and Company, New York 2002.

[3] F. Winterberg, in "Physics of High Energy Density", Preceedings of the International

School of Physics "Enrico Fermi", 14-26 July 1969, p.395-397, Academic Press New York, 1971.

[4] F.Winterberg, Raumfahrtforschung **15**, 208-217 (1971).

[5] F. J. Dyson, Physics Today **21**, 10, 41-45 (1968).

[6] Project Daedalus, A. Bond, A.R.Martin et al., J. British Interplanetary Society, Supplement, 1978.

[7] F.Winterberg, Phys. Rev. **174**, 212 (1968).

[8] F.Winterberg, Phys. Plasma **7**, 2654 (2000).

[9] F.Winterberg, J. Fusion Energy **2**, 377 (1982).

[10] F.Winterberg, "Can a Laser Beam ignite a Hydrogen Bomb?", United States Atomic Energy Commission, January 27, 1970 classified; declassified, July 11, 2007, $S - RD - 1$ $(NP - 18252)$.

[11] J. D. Balcomb et al. "Nuclear Pulse Space Propulsion Systems," Los Alamos Scientific Laboratory, classified November 1970, declassified July 10, 1979, LA - 4541 - MS.

[12] G. S. Janes, R. H. Levy, H. A. Bethe and B. T. Feld, Phys. Rev. 145, 925 (1966).